\documentclass[floats,floatfix,showpacs,amssymb,prd,twocolumn,superscriptaddress,nofootinbib,reprint]{revtex4-2}

\usepackage{amssymb,amsmath,verbatim,mathtools,needspace,enumitem,etoolbox,graphicx,physics,microtype,afterpage,bigints,textcomp,gensymb,tabularx,lmodern}
\usepackage[dvipsnames,table, usenames]{xcolor}
\definecolor{lightgray}{gray}{0.9}
\definecolor{lightblue}{rgb}{0.93,0.95,1.0}
\definecolor{linkcolor}{rgb}{0.846416, 0.272473, 0.421631}
\usepackage[unicode, colorlinks=true, linkcolor=linkcolor, citecolor=linkcolor, filecolor=linkcolor,urlcolor=linkcolor, pdfusetitle]{hyperref}
\usepackage[all]{hypcap}
\usepackage[T1]{fontenc}
\usepackage[utf8]{inputenc}
\usepackage{siunitx}
\usepackage{enumitem}
\usepackage[normalem]{ulem}
\usepackage{bm}
\usepackage{booktabs}
\usepackage{subcaption}

\newcommand\orcid[1]{\href{https://orcid.org/#1}{$\;\!$\includegraphics[width=0.7em]{orcid.png}}}

\definecolor{dodgerblue}{HTML}{1E90FF}
\definecolor{viennared}{HTML}{DA0A14}
\definecolor{ctorange}{HTML}{FF6C0C}
\definecolor{dkgreen}{HTML}{85CB33}

\usepackage{soul}

\newcommand{\bham}{
\affiliation{Institute for Gravitational Wave Astronomy
\& School of Physics and Astronomy, University of
Birmingham, Birmingham, B15 2TT, United Kingdom}}

\def\apj{\rm{ApJ}}
\def\apjl{\rm{ApJ}}

\def\aapr{\rm{A\&A~Rev.}}

\def\mnras{\rm{MNRAS}}
\def\prd{\rm{PRD}}

\hypersetup{
     colorlinks=true,
     linkcolor=dodgerblue,
     filecolor=dodgerblue,
     citecolor = dodgerblue,      
     urlcolor=dodgerblue,
     pdfauthor={Name}
     }

\begin{document}

\title{On the LISA science performance in observations of short-lived signals from massive black hole binary coalescences}
\author{Geraint Pratten, Antoine Klein, Christopher J. Moore, Hannah Middleton, Nathan Steinle, Patricia Schmidt and Alberto Vecchio}
\bham
\date{\today}

\begin{abstract}

The observation of massive black hole binary systems is one of the main science objectives of the Laser Interferometer Space Antenna (LISA). The instrument's design requirements have recently been revised: they set a requirement at $0.1\,\mathrm{mHz}$, with no additional explicit requirements at lower frequencies. This has implications for observations of the short-lived signals produced by the coalescence of massive and high-redshift binaries, one of the main science targets of the mission. Here we consider the most pessimistic scenario: the (unlikely) case in which LISA has no sensitivity below $0.1\,\mathrm{mHz}$.
We show that the presence of higher order modes (above the dominant $\ell = |m| = 2$ one) in the gravitational radiation from these systems, which will be detectable with a total signal-to-noise ratio $\sim 10^3$, allows LISA to retain the capability to accurately measure the physical parameters, the redshift, and to constrain the sky location. To illustrate this point, we consider a few selected binaries in the total (redshifted) mass range $4 \times10^6 - 4 \times 10^7\,M_\odot$ whose ($\ell = |m| = 2$) gravitational signals last between $\approx 12$ hours and $\approx 20$ days in band. We model the radiation using the highly accurate (spin-aligned) waveform approximant \texttt{IMRPhenomXHM} and carry out a fully coherent Bayesian analysis on the LISA noise-orthogonal time-delay-interferometry channels.

\end{abstract}

\maketitle

\section{Introduction}
\label{s:intro}

Reconstructing the merger history of massive ($\sim 10^7 - 10^5\,M_\odot$) black holes (MBHs), understanding their hosts and how they (co-)evolve over cosmic time, and determining their mass function and how it relates to the galaxies that harbour them are some of the most important open problems in astrophysics and cosmology, see \textit{e.g}~\cite{2010A&ARv..18..279V} for a review, and references therein. The Laser Interferometer Space Antenna (LISA)~\cite{lisa}, a gravitational-wave (GW) observatory which is currently being developed for science operation in the next decade, will provide major new observational capabilities to tackle these questions, see \textit{e.g.}~\cite{2022arXiv220306016A} and references therein. The results of this first GW survey capable of discovering MBH binaries throughout the Universe, complemented by a slew of other surveys and GW-triggered observations, \textit{e.g.} in optical (LSST~\cite{LSST}), mm/radio (ALMA~\cite{ALMA}/SKA~\cite{SKA}), near infra-red (JWST~\cite{JWST}) and X-ray (e.g. Athena~\cite{Athena+}), is likely to provide a significant leap forward in our understanding of the formation and evolution of entire populations of MBHs. 

LISA will observe the last hours-to-years of the lifetimes of MBH binaries. Observationally, the MBH binary merger rate is \textit{de-facto} unconstrained. Theoretical models predict wildly different rates depending on the assumptions, see \textit{e.g.} \cite{2021FrASS...8....7S}. As a consequence, the expected number of LISA detections varies from ${\cal O}(1)$ to ${\cal O}(100)$ per year with a mission lifetime of 4.5 years and a possible extension of up to 10 years. Regardless of the details of the model, the general trend is that one expects a handful of massive binaries ($> 10^6\,M_\odot$), as these constitute the bulk of the currently observed MBH population \cite{2015ApJ...813...82R,2018ApJ...852..131B,2019MNRAS.487.3404B}. In such a scenario, the detection of MBHs would be rare and from massive systems. If, however, comparably lighter MBHs are present, then the event rate could reach hundreds of binaries. 

The low-frequency sensitivity of LISA plays a vital role in determining the instrument's capability for detecting and characterising MBH binaries. In particular, it enters the analysis in two crucial ways that collectively determines the LISA science performance for the MBH parameter space.  

Firstly, it sets the maximum mass and redshift of the binaries that can be observed. For reference, the frequency associated to the dominant $\ell = |m|=2$ mode of gravitational radiation for a quasi-circular binary of non-spinning MBHs with individual (source-frame) masses%
\footnote{
Throughout the paper we adopt the convention that mass parameters in the source-frame are label by the suffix ``src'', so that the redshifted mass parameter, $m$ is related to its source-frame value, $m^{(\mathrm{src})}$, by $m = m^{(\mathrm{src})} (1 +z)$, where $z$ is the redshift.
}
$m_{1,2}^{(\mathrm{src})}$ and total mass $M^{(\mathrm{src})} = m_{1}^{(\mathrm{src})} + m_{2}^{(\mathrm{src})}$ at redshift $z$ is approximately
\begin{equation}
    f_\mathrm{ISCO}^{(2,2)} \sim 0.4\,\left[\frac{M^{(\mathrm{src})}\,(1 + z)}{10^7\,M_\odot}\right]^{-1}\,\mathrm{mHz}\,.
\end{equation}
Similarly, the frequency of the dominant quasi-normal mode of the remnant black hole produced in a merger as it settles to a quiescent state is approximately 
\begin{equation}
    f_\mathrm{ring}^{(2,2)} \sim 3.2\,\left[\frac{M^{(\mathrm{src})}\,(1 + z)}{10^7\,M_\odot}\right]^{-1}\,\mathrm{mHz}\,.
\end{equation}
These frequencies obey a natural hierarchy $f_{\rm ring} > f_{\rm ISCO}$, such that for a given instrument low-frequency cut-off, $f_\mathrm{low}$, there is a regime in which the mass and redshift of the binary are sufficiently high that the GW signal is pushed out of band, rendering the binary un-observable. 

Secondly, the low-frequency cut-off determines how long the signal will be in the detection band. The leading-order post-Newtonian duration of the ($\ell = |m| = 2$) signal from a binary in the LISA band can be written as 
\begin{align}
	\tau &\sim \left(\frac{3}{4 \eta} \right) \,\left(\frac{f_\mathrm{low}}{0.1\,\mathrm{mHz}}\right)^{-8/3}\, 
	\left[\frac{M^{(\mathrm{src})}\, (1 + z)}{10^7\,M_\odot}\right]^{-5/3}\,
	\mathrm{days},
	\label{eq:tau}
\end{align}
where $\eta = m_1 m_2/M^2$ is the symmetric mass ratio. The number of GW cycles in band scales as $N \sim \tau f_\mathrm{low}$, meaning that a shorter time in band will reduce our ability to accurately measure the physical parameters of the binary. In addition, 
this timescale has to be compared to $T_\mathrm{LISA} = 1\,\mathrm{yr}$, the time taken for the constellation to orbit the Sun and precess around the perpendicular to the ecliptic. The combination of these motions induces sky-location dependent modulations in the observed GW signal, providing the angular resolution of the LISA observatory. If $\tau \ll T_\mathrm{LISA}$, the observatory cannot resolve the source position in the sky, unless sky-location dependent information is encoded in the gravitational-wave polarization amplitudes.

A recent important development in the mission design is to impose a 
requirement at $0.1\,\mathrm{mHz}$, but no explicit sensitivity requirement below this frequency \cite{SciRD, 2021arXiv210801167B}. One may therefore wonder what would be the impact of the (unlikely) case in which sensitivity is lost below $f_\mathrm{low} = 0.1\,\mathrm{mHz}$. This means that as well as losing sensitivity to some of the most massive systems, reduced duration of MBHB signals in band impacts the science performance of the mission, particularly the ability to locate a source in the sky, which depends on the motion of the constellation as the coalescence takes place.

Many studies have been carried out over the years to explore LISA performance in observations of MBHBs, using different assumptions and/or approximations, see \textit{e.g.}~\cite{2009CQGra..26i4027A, 2016PhRvD..93b4003K, 2020PhRvD.102h4056M,2021PhRvD.103h3011M, 2020PhRvD.101h4053B, 2022PhRvD.105d4055K, 2022arXiv220710678M, 2022arXiv220813351W} and references therein. However, all these studies assumed a detection bandwidth that extended below $0.1\,\mathrm{mHz}$. 

In this paper, we take a worst case scenario approach concerning the impact of the new low-frequency design requirements, and consider the most pessimistic circumstance in which
LISA has no sensitivity below $f_\mathrm{low} = 0.1\,\mathrm{mHz}$. We explore the concomitant impact on the science capability of LISA in observing MBHBs by considering a small number of high mass systems over a total (redshifted) mass range $4 \times10^6 \,M_\odot - 4 \times 10^7\,M_\odot$. These systems produce short-lived coalescences, such that the dominant $\ell = |m| = 2$ harmonic is in the LISA band for $\approx 12\,\mathrm{hrs}$ for the heaviest binary through to $\approx 3\,\mathrm{weeks}$ for the lightest. We use a fully Bayesian analysis framework on the three time-delay-interferometry (TDI) LISA observables to compute the posterior probability density functions (PDF) of the source parameters. The gravitational-wave signal is modeled using the \texttt{IMRPhenomXHM} approximant, which is extremely accurate for binaries with a mass-ratio $q = m_2^{(src)}/m_1^{(src)} > 1/20$ and for BHs with spins (anti-)aligned with the orbital angular momentum. 

We demonstrate that LISA retains excellent performance in measuring the masses, spins, redshift (assuming a fixed $\Lambda$CDM cosmology), and sky localization of MBH binaries. This is due to the fact that these binaries will be observed with total SNRs $\sim 10^3$ and that multipoles beyond the dominant $(2,2)$-mode in the gravitational-wave strain will be detectable with SNRs $\sim 1-100$. The higher multipoles provide location-dependent information which is intrinsic to the gravitational-wave signal emitted by the binary and does not rely on the instrument's motion. In addition, the presence of higher multipoles also reduces correlations across many parameters, in agreement with~\cite{1994PhRvD..49.2658C, Varma:2014jxa, Varma:2016dnf, Chatziioannou:2019dsz, Shaik:2019dym, 2020PhRvD.101h4053B, 2021PhRvD.103h3011M, LIGOScientific:2020stg, 2022PhRvD.105d4055K, 2022arXiv220710678M, 2022arXiv220813351W}.  

The paper is organised as follows: in Sec.~\ref{s:methods} we describe the method and assumptions used in this work. Sec.~\ref{s:results} presents the main results, with further details provided in the Appendix. Sec.~\ref{s:concl} contains our conclusions and highlights future work.

\section{Method}
\label{s:methods}

Our goal is to explore the accuracy to which the parameters, $\bm{\theta}$, that describe a MBHB can be measured by LISA. In order to do this, we compute the posterior probability density function (PDF)
\begin{equation}
    p(\bm{\theta} | d) \propto {\mathcal L}\left(d | \bm{\theta}\right) p(\bm{\theta})
    \label{eq:posterior}
\end{equation}
given the data $d$. In the above expression, ${\mathcal L}\left(d | \bm{\theta}\right)$ is the likelihood, and $p(\bm{\theta})$ the prior, which we describe below.
%
%
\begin{table*}[ht!]
\caption{
Main properties of the injected sources and the recovered parameters. The injection values are denoted with the subscript ``inj''. The second column (HM) shows whether the injection and recovery are done including all the available modes in the waveform approximant (denoted with ``$\checkmark$") or just the $(2,2)-$mode (denoted with ``$\times$"). We stress that the approximant used in the injection and the likelihood are identical.
The masses in the table are the redshifted masses and the injected spins are $\chi_1 = 0.4$ and $\chi_2 = 0.2$. The redshift for all binaries is $z = 3$. For the recovered parameters, we show the median posterior value and the 90\% probability interval. $\Omega_\mathrm{90}$ is the 90\% probability interval of the 2D source location in the sky injection and $t_c$ is the time of coalescence with respect to an arbitrary reference epoch (the same applies to the phase of coalescence). The extrinsic parameters that are not reported in the table were the same for all the sources, with the following values: ecliptic longitude $l = 2.0$ and latitude $\sin b = 0.3$, inclination angle $\cos \iota = 0.9$, and polarisation $\psi = 0.4$ (the time and phase at coalescence are subject to an arbitrary choice of their zero-value, therefore we do not report them here). The binaries marked with a $\star$ indicate that the posteriors for $\Omega_{90}$ and $\Delta t_c$ are multimodal (MM) and that due caution should be used in interpreting these numbers.
}
\vspace{0.1cm}
\def\arraystretch{1.5}
\centering
\begin{tabular}{l|c|cc|ccccc|ccccc|c}
\toprule
ID                       &
HM                    &
$m_1^{\rm inj}$                           &  
$m_2^{\rm inj}$                           &
$f_\mathrm{low}$                &
SNR                             &
$\tau_{(\ell = 2)}$             &
$\tau_{(\ell = 3)}$             &
$\tau_{(\ell = 4)}$             &
$m_1$                    &
$m_2$                    &
$z$                      & 
$\Omega_\mathrm{90}$                 &
$\Delta t_c$             &       
MM               \\
                                &
                                &
$\left[10^6 M_{\odot}\right]$        &  
$\left[10^6 M_{\odot}\right]$        &
$[\mathrm{mHz}]$                &
                                &
$[\mathrm{days}]$                  &
$[\mathrm{days}]$                  &
$[\mathrm{days}]$                  &
$\left[ 10^6 M_{\odot}\right]$        & 
$\left[ 10^6 M_{\odot}\right]$        & 
                                &
$\left[ \rm{deg}^2 \right]$     &
$[\mathrm{s}]$                  & 
\\
\hline
\hline
I                 &
$\checkmark$      &
$32.0$  &
$8.0$  &
0.1               &
606               &
$0.46$            &
$1.15$            &
$2.63$            &
${32.00}_{-0.07}^{+0.08}$               &
${7.99}_{-0.18}^{+0.18}$               &
${3.00}_{-0.05}^{+0.06}$               & 
$527$            &
${-3}_{-56}^{+440}$              & 
$\star$
\\
Ia                &
$\checkmark$          &
$32.0$  &
$8.0$  &
0.05              &
630               &
$2.95$            &
$8.29$                  &
$18.4$                  &
${32.00}_{-0.07}^{+0.07}$               &
${7.97}_{-0.09}^{+0.10}$              &
${3.00}_{-0.03}^{+0.03}$              & 
$599$           &
${5}_{-225}^{+632}$               & 
$\star$
\\
Ib                &
$\checkmark$          &
$32.0$  &
$8.0$  &
0.01              &
655               &
$215$          &
$654$          &
$1409$                  &
${32.01}_{-0.09}^{+0.08}$               &
${7.96}_{-0.05}^{+0.05}$              &
${2.99}_{-0.02}^{+0.02}$               & 
$0.4$           &
${1.6}_{-22.7}^{+22.1}$               & 
\\
I                 &
$\times$          &
$32.0$  &
$8.0$  &
0.1               &
550               &
$0.46$            &
$-$               &
$-$               &
${31.94}_{-0.41}^{+0.53}$               &
${8.05}_{-0.58}^{+0.54}$               &
${2.66}_{-0.69}^{+0.55}$               & 
$4459$            &
${-52}_{-243}^{+712}$               &
$\star$
\\
\hline
II                &
$\checkmark$          &
$16.0$  &
$4.0$  &
0.1               &
1159              &
$1.31$            &
$4.35$                  &
$9.36$                  &
${16.00}_{-0.02}^{+0.02}$               &
${4.01}_{-0.02}^{+0.02}$               &
${3.01}_{-0.01}^{+0.01}$               & 
$5.2$           &
${2}_{-5}^{+428}$               &
$\star$
\\
II &
$\times$          &
$16.0$  &
$4.0$  &
0.1               &
1037              &
$1.31$            &
$-$               &
$-$               &
${15.98}_{-0.09}^{+0.11}$               &
${4.03}_{-0.13}^{+0.05}$              &
${2.73}_{-0.17}^{+0.26}$              & 
$1986$             &
${16}_{-319}^{+432}$               &
$\star$ \\
\hline
III               &
$\checkmark$          &
$6.4$  &
$1.6$  &
0.1               &
1864              &
$6.80$            &
$20.0$               &
$43.1$               &
${6.40}_{-0.01}^{+0.01}$               &
${1.60}_{-0.004}^{+0.004}$               &
${3.00}_{-0.01}^{+0.01}$              & 
$1.1$          &
${-0.3}_{-1.7}^{+1.8}$               &
\\
III &
$\times$          &
$6.4$  &
$1.6$  &
0.1               &
1814              &
$6.80$            &
$-$               &
$-$               &
${6.40}_{-0.01}^{+0.01}$              &
${1.60}_{-0.004}^{+0.004}$               &
${2.72}_{-0.23}^{+0.39}$             & 
$36.1$            &
${1.1}_{-21.9}^{+11.1}$              &
\\
\hline
IV                &
$\checkmark$          &
$3.2$  &
$0.8$&
0.1               &
1933              &
$22.1$           &
$63.6$               &
$137$               &
${3.20}_{-0.005}^{+0.005}$               &
${0.80}_{-0.001}^{+0.001}$               &
${3.00}_{-0.01}^{+0.01}$               & 
$0.1$           &
${-0.01}_{-0.94}^{+0.96}$              &
\\
IV                &
$\times$          &
$3.2$  &
$0.8$&
0.1               &
1931              &
$22.1$           &
$-$               &
$-$               &
${3.20}_{-0.004}^{+0.004}$              &
${0.80}_{-0.001}^{+0.001}$              &
${2.91}_{-0.29}^{+0.27}$               & 
$11.5$            &
${1.4}_{-18.1}^{+7.5}$               &
\\
\bottomrule
\end{tabular}
\label{tab:injections}
\end{table*}
%
%

We employ a coherent analysis using the three noise-orthogonal TDI channels, $\lbrace A, E, T \rbrace$, which constitute the data $d$ in Eq.~(\ref{eq:posterior}) and are synthesised from the LISA phase-meters~\cite{2002PhRvD..66l2002P}. The LISA response is modelled following the rigid adiabatic approximation, \textit{e.g.} Ref.~\cite{2004PhRvD..69h2003R}, implying that we are working with TDI variables of generation $1.5$. The actual analysis of the data will require a more sophisticated TDI scheme to suppress the laser frequency noise but for the purpose of this paper the approximation does not affect the core results. The likelihood in Eq.~(\ref{eq:posterior}) can be written as~\cite{1994PhRvD..49.2658C}
\begin{align}
\ln {\mathcal L}(d | \bm{\theta}) &= -\!\sum_{k} \frac{\langle d_k - h_k(\bm{\theta}) | d_k - h_k(\bm{\theta}) \rangle_k}{2}\! + \mathrm{const},
\label{eq:like}
\end{align}
where the sum is taken over the three TDI channels, labelled by $k$, and $h_k$ denotes the TDI signal produced by a MBH binary with source parameters $\bm{\theta}$. The noise-weighted inner-product is defined in the usual way
\begin{align}
    \langle a | b \rangle_k &= 2 
    \int_{f_\mathrm{low}}^{f_\mathrm{high}(\bm{\theta})}\!
    \frac{\tilde a(f) \tilde b^*(f) + \tilde a^*(f) \tilde b(f)}{S_k(f)} \, \mathrm{d}f \,.
    \label{eq:innerprod}
\end{align}
In the above equation, $\tilde a(f)$ denotes the Fourier transform of the time series $a(t)$,  $S_k(f)$ the noise power spectral density of the $k$-th TDI channel, and $f_\mathrm{low} = 0.1\,\mathrm{mHz}$ is LISA's low-frequency cut-off. The highest frequency of the signal produced by a MBH binary, $f_\mathrm{high}(\bm{\theta})$, depends on the source parameters. We use noise spectral densities $S_k(f)$ as given in ESA Science Requirements Document~\cite{SciRD, 2021arXiv210801167B}, with the important addition of a hard low-frequency cut-off, $f_\mathrm{low}$, at $0.1\,\mathrm{mHz}$. The unresolved galactic confusion noise is modeled according to the analytical fit given in~\cite{SciRD}. 

For the computation of the gravitational-wave polarizations, $h_{+,\times}$, we assume that, regardless of formation history, the binaries circularise by the time they enter the LISA sensitivity band due to radiation reaction, e.g.~\cite{Peters:1964zz}. We use \texttt{IMRPhenomXHM}~\cite{Pratten:2020fqn,Garcia-Quiros:2020qpx} to model the full coalescence of quasi-circular binaries, i.e. the inspiral, merger, and ringdown. The aligned-spin modes, $h_{\ell m}$, are calibrated to numerical relativity and include the $(\ell,|m|) = \lbrace (2,2),(2,1),(3,3),(3,2),(4,4)\rbrace$ multipoles. In this study we only consider aligned-spin binaries, \textit{i.e.} systems in which the MBH spins are parallel or anti-parallel to the orbital angular momentum. Spins that are mis-aligned with the orbital angular momenta, which induces precession of the orbital plane and the spins themselves, are not considered in this study, though we will return to them in a future study.

With these assumptions, the signal parameter set $\bm{\theta}$ consists of: two independent mass parameters, \textit{e.g.} the (redshifted) chirp mass $\mathcal{M}_c$ and mass ratio $q\equiv m_2/m_1 \le 1$, the dimensionless spins, $\chi_{i} \equiv \vec{S}_{1,2} \cdot \hat{L}/m_{1,2}^2$, where $\vec{S}_{i}$ is the intrinsic spin-angular-momenta of the BH and $\hat{L}$ is the (constant) orbital angular momentum unit vector, the location of the source in the sky, in terms of the ecliptic latitude $b$ and longitude $l$, the luminosity distance of the source $d_L(z)$, or equivalently the redshift $z$ (we assume standard $\Lambda$CDM cosmology according to Planck 2018~\cite{2020A&A...641A...6P}), two parameters that describe $\hat{L}$, taken to be the inclination angle $\iota$ with respect to the line-of-sight and the GW polarization phase $\psi$, and, finally, the coalescence time $t_c$ and associated GW phase $\phi_c$. Note that $t_c$ and $\phi_c$ correspond to a gauge-freedom and are defined with respect to an arbitrary reference value. 

To perform all parameter estimation, we use \texttt{Balrog}, a software package under development for LISA data analysis, see \textit{e.g.}~\cite{2020ApJ...894L..15R, 2021PhRvD.104d4065B, 2022arXiv220403423K, 2022arXiv221010812F}. We perform full Bayesian inference on simulated data using the nested sampling \cite{Skilling:2004} algorithm implemented in \texttt{dynesty} \cite{Speagle:2020} to evaluate Eq.~\eqref{eq:posterior}. All injections are performed in zero noise. 

\begin{figure*}[ht!]
    \centering
    \begin{subfigure}{0.9\textwidth}
    \includegraphics[width=\textwidth]{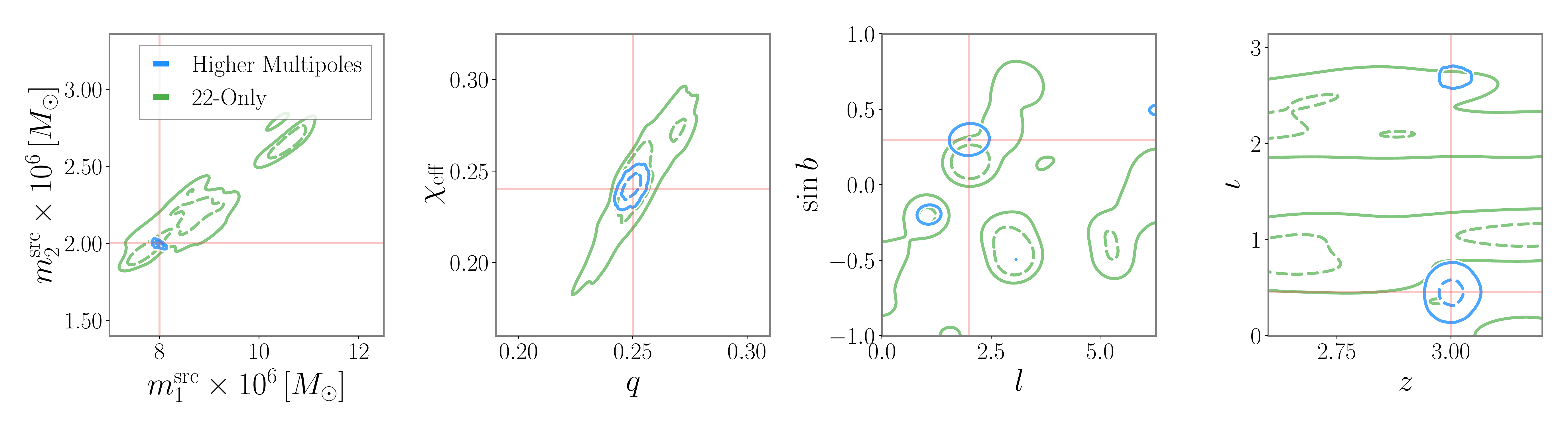}
        \caption{
        System ID I (See cornerplot in Fig.~\ref{fig:binary1_corner}).
        }
    \label{fig:binary1}
    \end{subfigure}

    \begin{subfigure}{0.9\textwidth}
    \includegraphics[width=\textwidth]{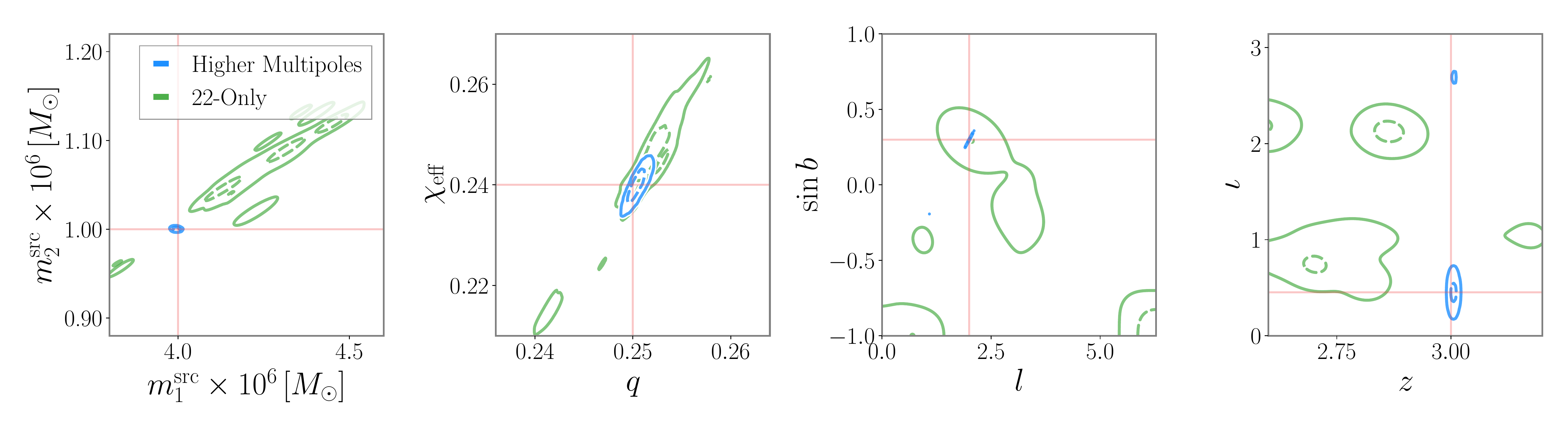}
        \caption{
        System ID II (See cornerplot in Fig.~\ref{fig:binary2_corner}).
        }
    \label{fig:binary2}
    \end{subfigure}

    \begin{subfigure}{0.9\textwidth}
    \includegraphics[width=\textwidth]{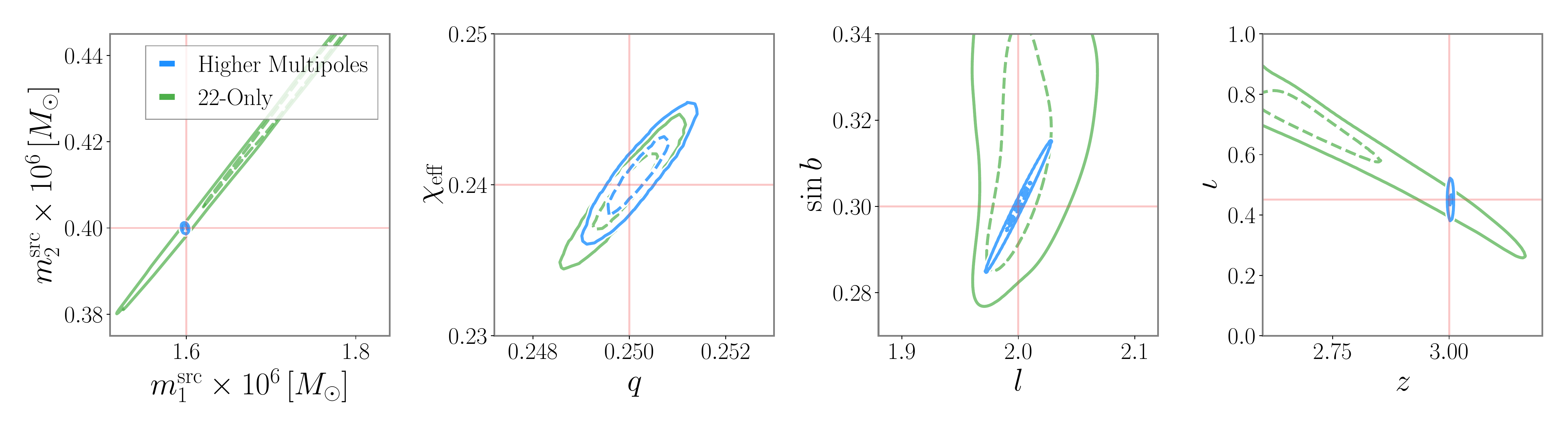}
        \caption{
        System ID III (See cornerplot in Fig.~\ref{fig:binary3_corner}).
        }
    \label{fig:binary3}
    \end{subfigure}

    \begin{subfigure}{0.9\textwidth}
    \includegraphics[width=\textwidth]{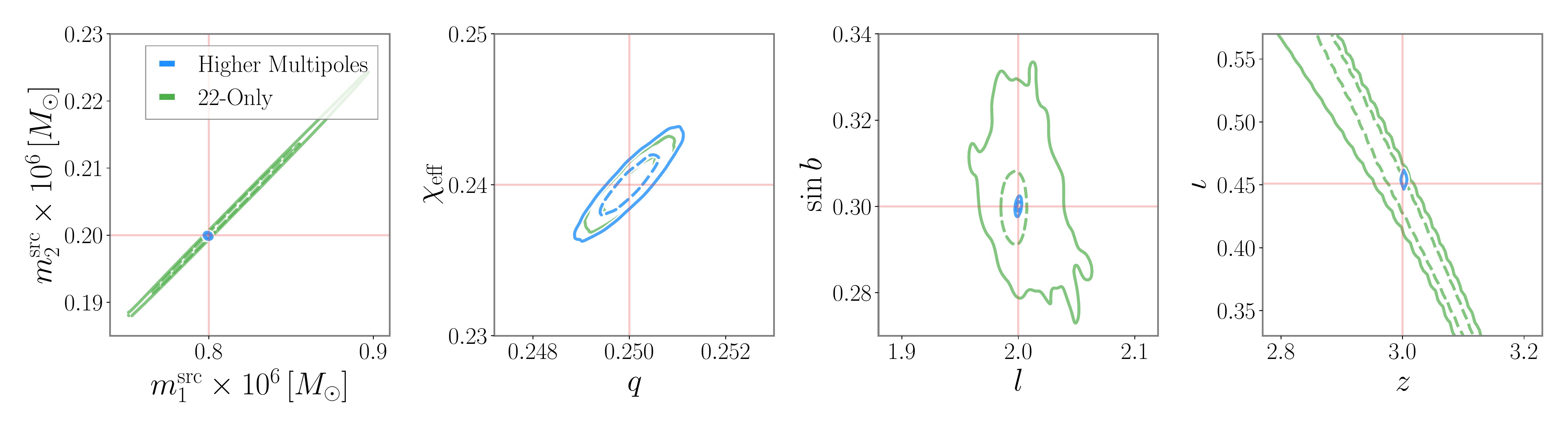}
        \caption{
        System ID IV (See cornerplot in Fig.~\ref{fig:binary4_corner}).
        }
    \label{fig:binary4}
    \end{subfigure}
\caption{Marginalised posterior density functions for selected parameters for the four systems of Table~\ref{tab:injections}, with $f_\mathrm{low} = 0.1\,\mathrm{mHz}$. The plots show the 50\% (dashed) and 90\% (solid) probability contours inferred when using higher multipoles (blue) and just the $(2,2)$-mode (green). The solid red lines denote the injection values. 
}
\end{figure*}

\section{Results}
\label{s:results}

\begin{figure*}[ht!]
    \includegraphics[width=\textwidth]{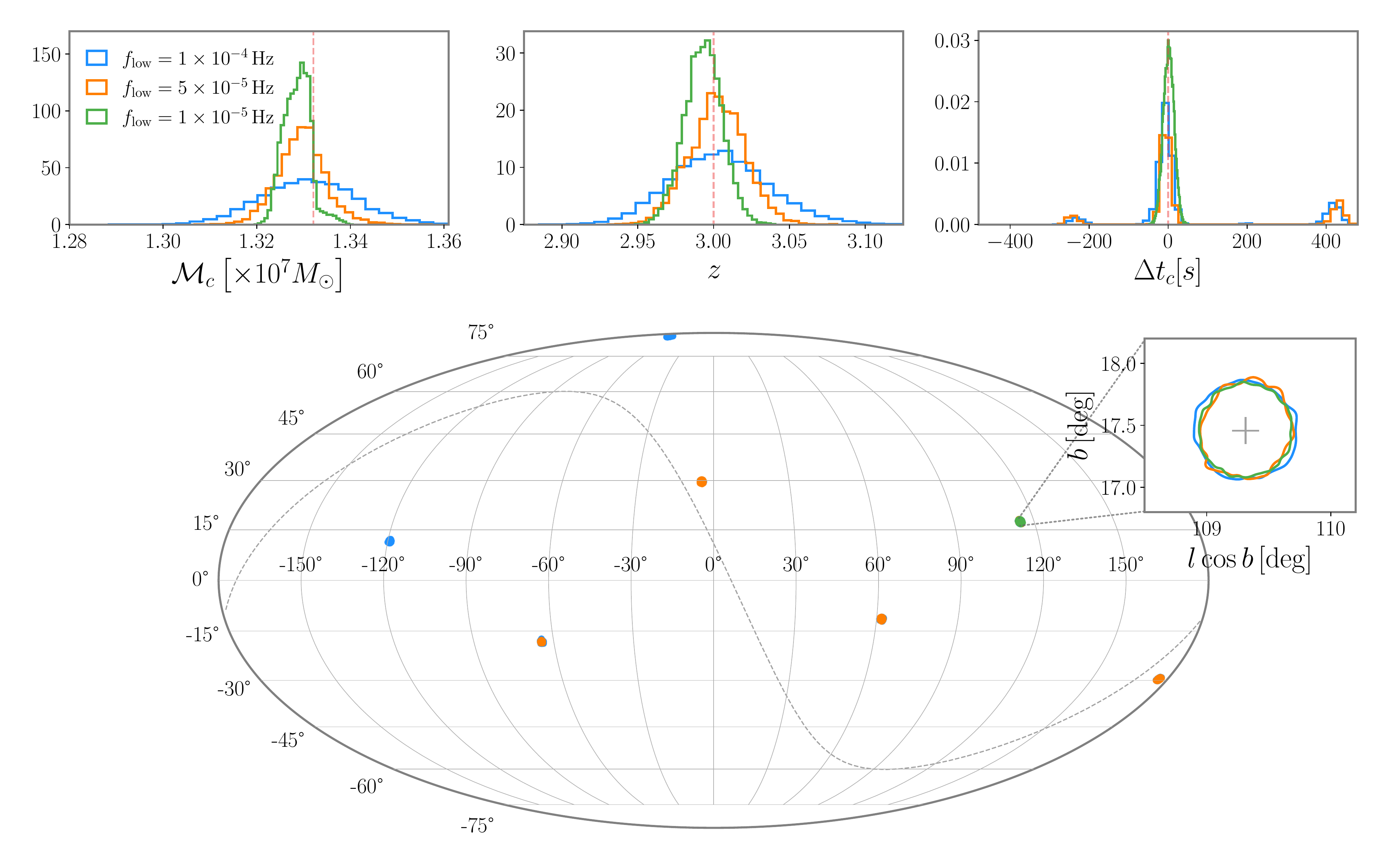}
    \caption{Impact of lowering the cutoff frequency for the LISA sensitivity band, $f_\mathrm{low}$, on the inferred parameters of a binary back hole with $m_1 = 3.2\times 10^7\,M_\odot$ and $m_1 = 8\times 10^6\,M_\odot$ (Run ID: I, Ia and Ib in Table~\ref{tab:injections}). Decreasing the cutoff frequency effectively means a longer duration signal in within LISA's sensitive band, typically resulting in parameter degeneracies being broken due to the amplitude and phase modulations induced by LISA's orbital motion. Waveform models incorporating higher multipoles can partially mitigate against such changes due to the dependence of the multipoles on the binary geometry, i.e. its sky location and orientation. The coalescence time $t_c$ is defined at the solar system barycenter; therefore different possible sky posteriors translate into different possible coalescence times.
}
    \label{fig:binary1_flow}
\end{figure*}

We consider a small number of representative systems for this study. The key source parameters are listed in Table~\ref{tab:injections}, and are chosen to be broadly consistent with predictions from theoretical models, see \textit{e.g.}~\cite{2016PhRvD..93b4003K, 2012MNRAS.423.2533B, 2022arXiv220306016A} and references therein. We focus on binaries whose combination of masses and redshift are mostly affected by the new low-frequency design requirement, and consider four MBHBs with redshifted total mass in the range $4 \times 10^6\,M_\odot - 4 \times 10^7\,M_\odot$ (with $z = 3$ in all cases). For all the systems we set the mass ratio to  $q = 1/4$, and the (aligned) spins to moderate values, $\chi_1 = 0.4$ and $\chi_1 = 0.2$, corresponding to an effective aligned spin $\chi_\mathrm{eff} = 0.24$, where \cite{Damour:2001tu,Racine:2008qv,Ajith:2009bn}
\begin{align}
    \chi_{\rm eff} &= \frac{(m_1 \vec{\chi}_1 + m_2 \vec{\chi}_2) \cdot \vec{L}_N}{M}.
    \label{eq:chi_eff}
\end{align}

As shown in Tab.~\ref{tab:injections}, for all these binaries the time spent in band by the dominant $\ell = |m| = 2$ mode of the gravitational radiation is much shorter than $T_\mathrm{LISA}$: it ranges from $\approx 0.5\,\mathrm{days}$ (for $M = 4\times 10^7\,M_\odot$) to $\approx 22\,\mathrm{days}$ (for $M = 4\times 10^6\,M_\odot$). However, gravitational radiation from higher ($\ell = 3$ and $4$) multipoles is within the observational band for longer periods, with radiation from each multipole $\ell$ in the sensitivity band for a time $\tau_\ell=\tau (\ell/2)^{8/3}$, where $\tau$ is approximated by Eq.~(\ref{eq:tau}). They range from $\approx 3\,\mathrm{days}$ for $M = 4\times 10^7\,M_\odot$ through to $\approx 137\,\mathrm{days}$ for $M = 4\times 10^6\,M_\odot$.
The higher multipoles also contribute a factor of a few through to several tens to the total SNR of the binary depending on the specific source parameters.

For each of the four choices of total mass (labeled ``ID'' in Table~\ref{tab:injections}) we generate synthetic data in which we set $f_\mathrm{low} = 0.1\,\mathrm{mHz}$ and inject and recover the coalescence signal using only the $(2,2)-$mode and the full set of modes available in the approximant \texttt{IMRPhenomXHM} (denoted with ``$\times$" and ``$\checkmark$'', respectively, in column ``HM" of Tab.~\ref{tab:injections}). We stress that for each of the runs, the injection and recovery is done with an \textit{identical} waveform model, as we want to explore the impact of the higher multipoles on the analysis and not to concern ourselves with parameter biases arising from systematic errors between waveform approximants. The total signal-to-noise ratio produced by these binaries is in the range $\approx 6\times 10^2 - 2\times 10^3$. 

To provide a quantitative assessment of the impact on the LISA science performance imposed by the new low-frequency design requirement, we repeat the analysis for the heaviest binary (ID ``I'' in Tab.~\ref{tab:injections}), $m_1 = 3.2\times 10^7\,M_\odot$ and $m_2 = 8\times 10^6\,M_{\odot}$, using a range of low-frequency cut-offs, 
\textit{i.e.} $f_\mathrm{low} = 0.05\,\mathrm{mHz}$ (``Ia'') and $f_\mathrm{low} = 0.01\,\mathrm{mHz}$ (``Ib'').

The results are summarised in Tab.~\ref{tab:injections}, and Fig.~\ref{fig:binary1}-\ref{fig:binary4} show the 2D PDFs for selected parameters of interest for each of the MBH binaries. We also compare the results obtained by considering only the $(2,2)$-mode against those obtained using the full range of higher multipoles available. Full cornerplots are provided in Fig.~\ref{fig:binary1_corner}-\ref{fig:binary4_corner} of the Appendix. In Fig.~\ref{fig:binary1_flow} we compare the posterior PDFs on selected parameters obtained by assuming different low-frequency cut-offs for the heaviest system, to provide a quantitative indication of the impact of the new design requirements. 

The first general trend observed is that one draws radically different conclusions on the LISA science performance, described by the size of the statistical errors on the system parameters, if one considers only the dominant $(2,2)$-mode (green contours in Fig.~\ref{fig:binary1}-\ref{fig:binary4}) or if one includes the full set of higher multipoles, confirming results in other portions of the LISA parameter space~\cite{2021PhRvD.103h3011M, 2020PhRvD.101h4053B, 2022PhRvD.105d4055K, 2022arXiv220710678M}. For the binaries considered here, the $90\%$ confidence intervals are typically larger by a factor $\sim 10-1000$ for the $(2,2)$-mode results with respect to the full multipolar results. The inclusion of higher multipoles has a strong impact on the correlations and degeneracies between parameters. Below, we focus on discussing the results obtained when using only the full set of modes.

We start by discussing the results on the physical parameters, \text{i.e.} the masses and spins. For the latter, we focus on the effective spin parameter $\chi_\mathrm{eff}$, as defined in Eq.~\eqref{eq:chi_eff}, given its relevance for aligned-spin waveforms.

LISA retains excellent capability, by most astronomical standards, to measure the \textit{individual} source-frame masses (assuming $\Lambda$CDM standard cosmology), which can be measured to $\approx 1 \%$ (or better) for MBHBs within the parameter range considered here. This comes from the combination of the measurement of the detector-frame masses and the redshift (for the latter see below). In fact, to understand the difference between the measurements of detector-frame vs source-frame masses it is useful to compare the results of Fig.~\ref{fig:binary1} --~\ref{fig:binary4}, where we plot source-frame masses, with those of Tab.~\ref{tab:injections}, where we report results for detector-frame masses. 

LISA equally maintains good capability to measure the effective spin parameter $\chi_\mathrm{eff}$, which is always measured with an error (at 90\% credibility) $\approx 0.1 - 0.01$. Due to the signatures of higher multipoles, the correlation between $\chi_\mathrm{eff}$ and the mass ratio $q$ is largely broken, see Fig.~\ref{fig:binary1} --~\ref{fig:binary4}, even for signals that span a very short period in the LISA sensitivity band. The general trend is that the error on $\chi_\mathrm{eff}$ decreases as the number of wave cycles increases, therefore it is smaller for the lighter binaries considered here, which also happen to have the larger SNRs, see also \cite{Vitale:2014mka,Purrer:2015nkh,Vitale:2016avz,Pratten:2020igi,Krishnendu:2021cyi}. 

We now turn our attention to the parameters that describe the location of the source in the sky, and the time at which the coalescence takes place. These are important to (eventually) infer the merger history of MBHBs across cosmic time, and relate GW-detected sources to their environment (galaxy and/or galaxy cluster) through observational campaigns in the electro-magnetic spectrum across many wavelengths~\cite{2020PhRvD.102h4056M, McGee:2018qwb, 2022arXiv220710678M}. 

The redshift determination, which comes from the direct measurement of the luminosity distance assuming a standard cosmology, is constrained to $z_{\rm inj} \pm 0.1$ for all the binaries considered here, where all systems have a fixed redshift $z=3$. 
This comes from a combination of the comparatively high SNR, an accurate measurement of the chirp mass, and the fact that the higher modes break correlations between parameters, in particular the degeneracy between the distance and the inclination angle \cite{1994PhRvD..49.2658C,Nissanke:2009kt,Schutz:2011tw,LIGOScientific:2017vwq,Chen:2018omi,Usman:2018imj}. The redshifts are measured with a 90\% credibile interval of $\Delta z \approx 0.1$ for the heaviest binary ($M = 4\times 10^7\,M_\odot$) through to $\Delta z \approx 0.01$ for the lightest ($M = 4\times 10^6\,M_\odot$).

One may expect that it would be impossible for LISA to identify the source location in the sky, due to the signals being in band for a time $\ll T_\mathrm{LISA}$. However, higher multipoles, which also encode information about the source geometry, play a particularly important role. For the shortest of the signals considered here, the dominant $\ell = |m| = 2$ mode is in band for half a day and the resulting 2D posterior PDF on the sky location is multi-modal, albeit with well constrained modes, whose local $90\%$ probability regions are $\sim 1\,\mathrm{deg}^2$. By the time the $\ell = |m| = 2$ mode is in band for $\approx$ one week (for binary III, the $\ell = 3$ and $\ell = 4$ multipoles are in band for 20 days and 43 days respectively) the binary location is well constrained to a single 
mode with 90\% probability interval, $\Omega_{90} \approx 1 \rm{deg}^2$. As the mass of the binary descreases and the inspiral time goes up, the angular resolution increases accordingly. 

The determination of the merger time is also affected by the multi-modality of the sky location. For the shortest of the signal, there are multiple well localised merger times that span an overall window of $\approx 10\,\mathrm{min}$. As the number of cycles in band increases, the multi-modality disappears. For the lightest binary considered here, the merger time is measured to within a 90\% confidence error of approximately $1\,\mathrm{sec}$.

\section{Conclusions}
\label{s:concl}

We have explored the LISA science performance for observations of short-lived, high-redshift MBH binaries in light of the new low-frequency LISA design requirement that does not explicitly stipulate any sensitivity at frequencies below $0.1\,\mathrm{mHz}$. We have considered a small number of binary parameters such that the signal is visible in band from $0.1\,\mathrm{mHz}$ through to merger and for a time significantly shorter than $T_\mathrm{LISA}$. 

The results of this study can be taken as an indication that LISA would preserve its core science capabilities (despite some loss of performance) for short-lived signals, even if $0.1\,\mathrm{mHz}$ was a hard, low-frequency cutoff. However, this work should not be regarded as a detailed characterisation of the expected LISA science performance: statistical errors that characterise the measurement accuracy often strongly depend on the true binary parameters. Hence, a characterisation of the science performance would require a large, and computationally costly, Bayesian inference campaign that goes beyond the scope of this paper.

As a corollary of this work, we have shown that including only the dominant $\ell = |m| = 2$ modes results in forecasting the accuracy of parameter determination which is worse by a factor $\sim 10-1000$ for the very short-lived signals considered here, depending on the specific parameter under consideration and the binary parameters. It is well known that higher modes play an important role in LISA observations for MBHBs that produce SNRs $\sim 10^2-10^3$; however the few studies carried out so far~\cite{2020PhRvD.102b3033K, 2021PhRvD.103h3011M, 2022PhRvD.105d4055K, 2022arXiv220813351W, 2022arXiv220710678M} have mainly considered lighter binaries that are observable for months-to-years and/or used a low-frequency cutoff $\approx 0.01\,\mathrm{mHz}$, which radically change the observation.

Various assumptions made in this work should be relaxed in the future in the context of further systematic studies. Our assumption of aligned-spin systems has been driven by computational efficiency and is not necessarily supported by astrophysical expectations, see e.g.~\cite{2007ApJ...661L.147B,2022arXiv221100044S} and reference therein. 
If spins are mis-aligned with respect to the orbital plane, additional phase and amplitude modulations induced by precession \cite{Apostolatos:1994mx,Schmidt:2012rh} will couple to LISA's ability to extract essentially the whole set of parameters of a binary and may actually \textit{improve} LISA measurements~\cite{2004PhRvD..70d2001V, 2006PhRvD..74l2001L}. We have also assumed that the power-spectral density of the noise is known. This is not going to be the case for real observations, and accounting for additional elements of uncertainty, including data gaps, non-stationary noise, and transient instrumental glitches, will be required for more realistic predictions \cite{Baghi:2019eqo,Dey:2021dem, Edwards:2020tlp}.

\section*{Acknowledgements}

We thank Antoine Petiteau for useful discussions. G.P. is grateful for support from a Royal Society University Research Fellowship URF{\textbackslash}R1{\textbackslash}221500 and STFC grant ST/V005677/1. A.K., H.M., C.J.M. and A.V. acknowledge the support of the UK Space Agency, Grant No. ST/V002813/1. N.S. is supported by Leverhulme Trust Grant No. RPG-2019-350, European Union's H2020 ERC Starting Grant No. 945155--GWmining, and Cariplo Foundation Grant No. 2021-0555. P.S. acknowledges support from STFC grant ST/V005677/1. A.V. acknowledges the support of the Royal Society and Wolfson Foundation. Computational resources used for this work were provided by the University of Birmingham’s BlueBEAR High Performance Computing facility and the Bondi HPC Cluster at the Birmingham Institute for Gravitational Wave Astronomy.

\section{Appendix}
\label{s:appendix}

Here we provide full corner-plots of the posterior probability density functions on the system parameters for the runs listed in Table~\ref{tab:injections} and discussed in the main text. For the sake of readability, we omit the coalescence phase and the polarization, which do not carry important physical information about the binaries.

\begin{figure*}[ht!]
    \centering
    \includegraphics[width=\textwidth]{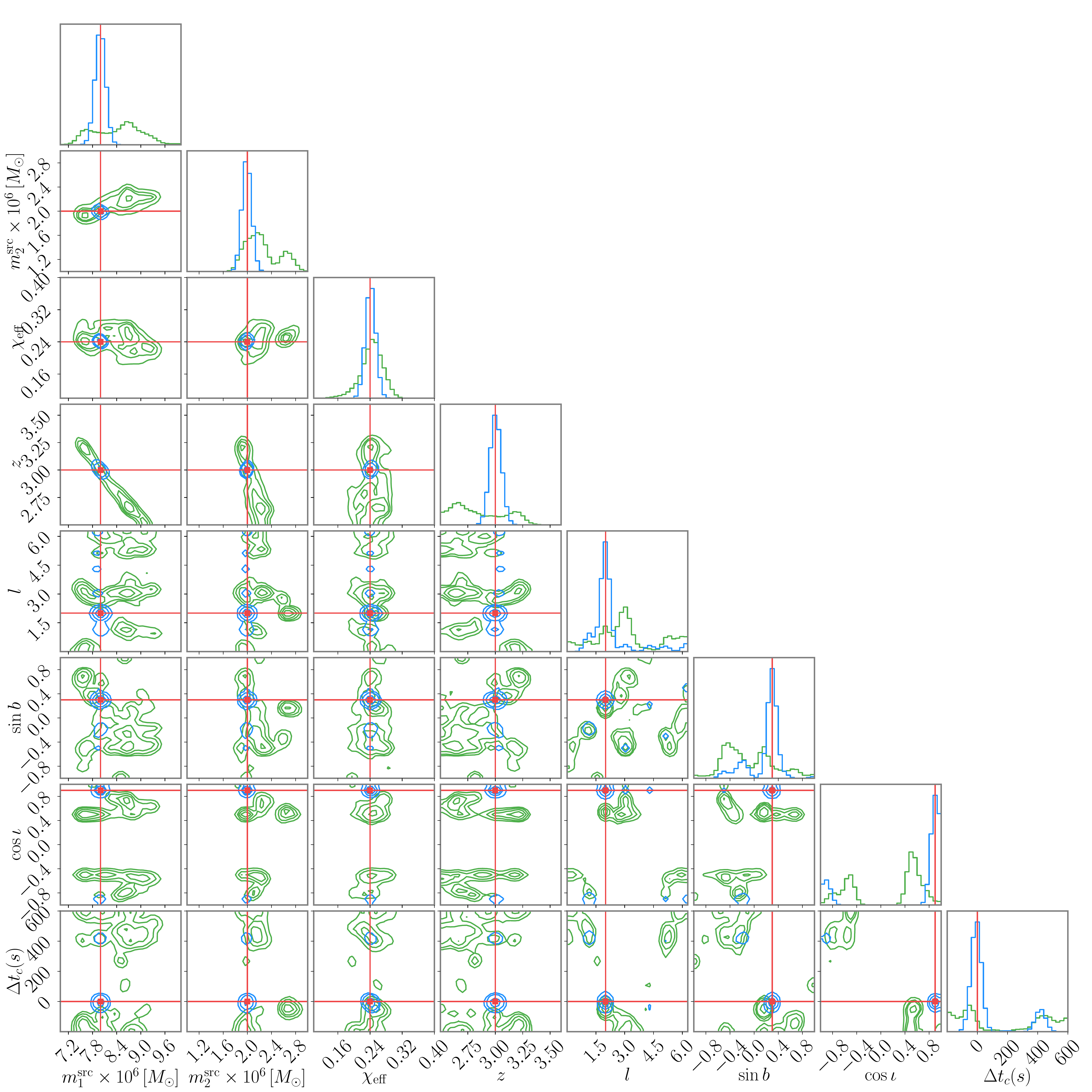}
        \caption{
        Binary I. Run with higher multipoles (blue) and with the $(2,2)$-mode only (green). The red lines denote the injected values. }
        \label{fig:binary1_corner}
\end{figure*}

\begin{figure*}[ht!]
    \centering
    \includegraphics[width=\textwidth]{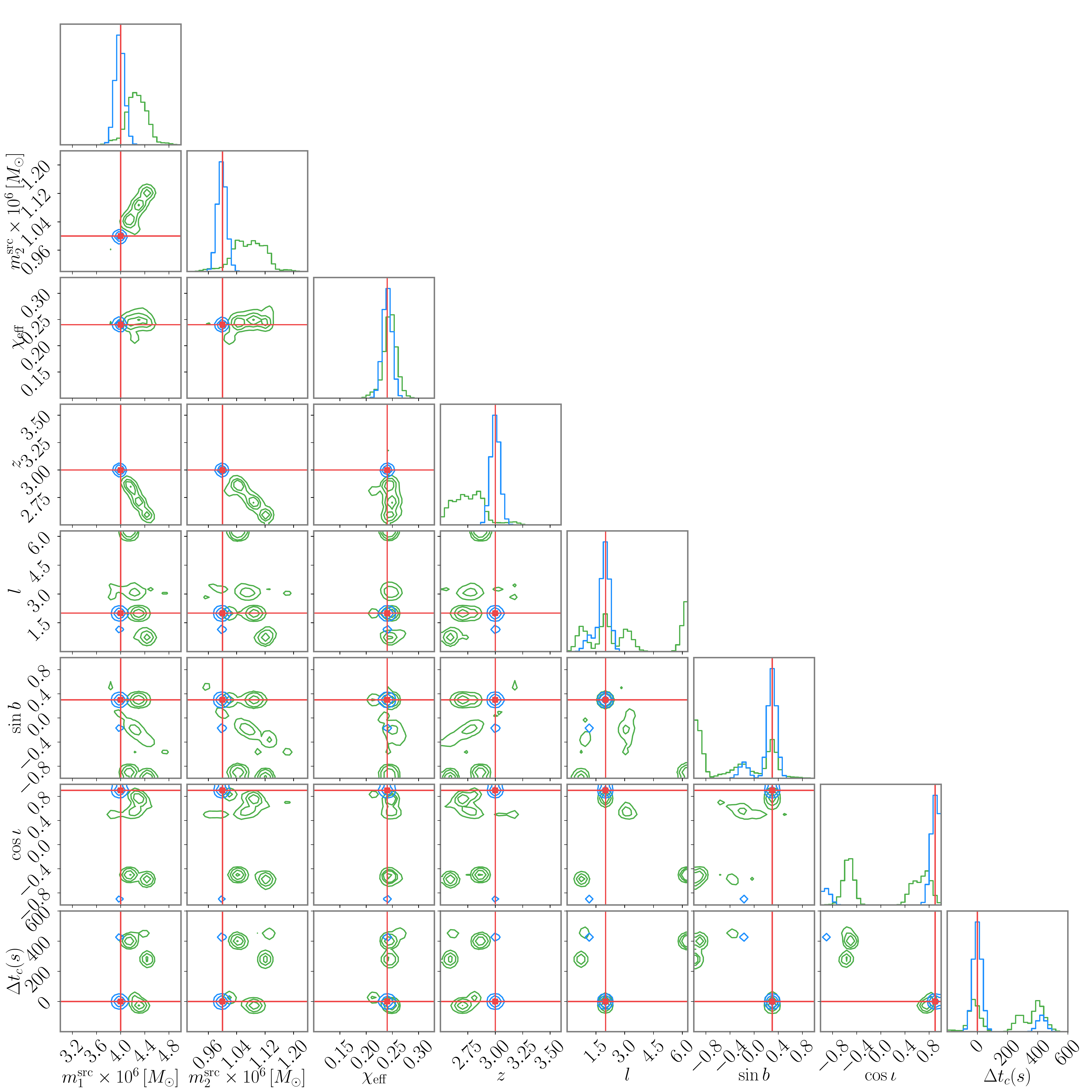}
        \caption{
        Binary II. Run with higher multipoles (blue) and with the $(2,2)$-mode only (green). The red lines denote the injected values. 
        }
        \label{fig:binary2_corner}
\end{figure*}

\begin{figure*}[ht!]
    \centering
    \includegraphics[width=\textwidth]{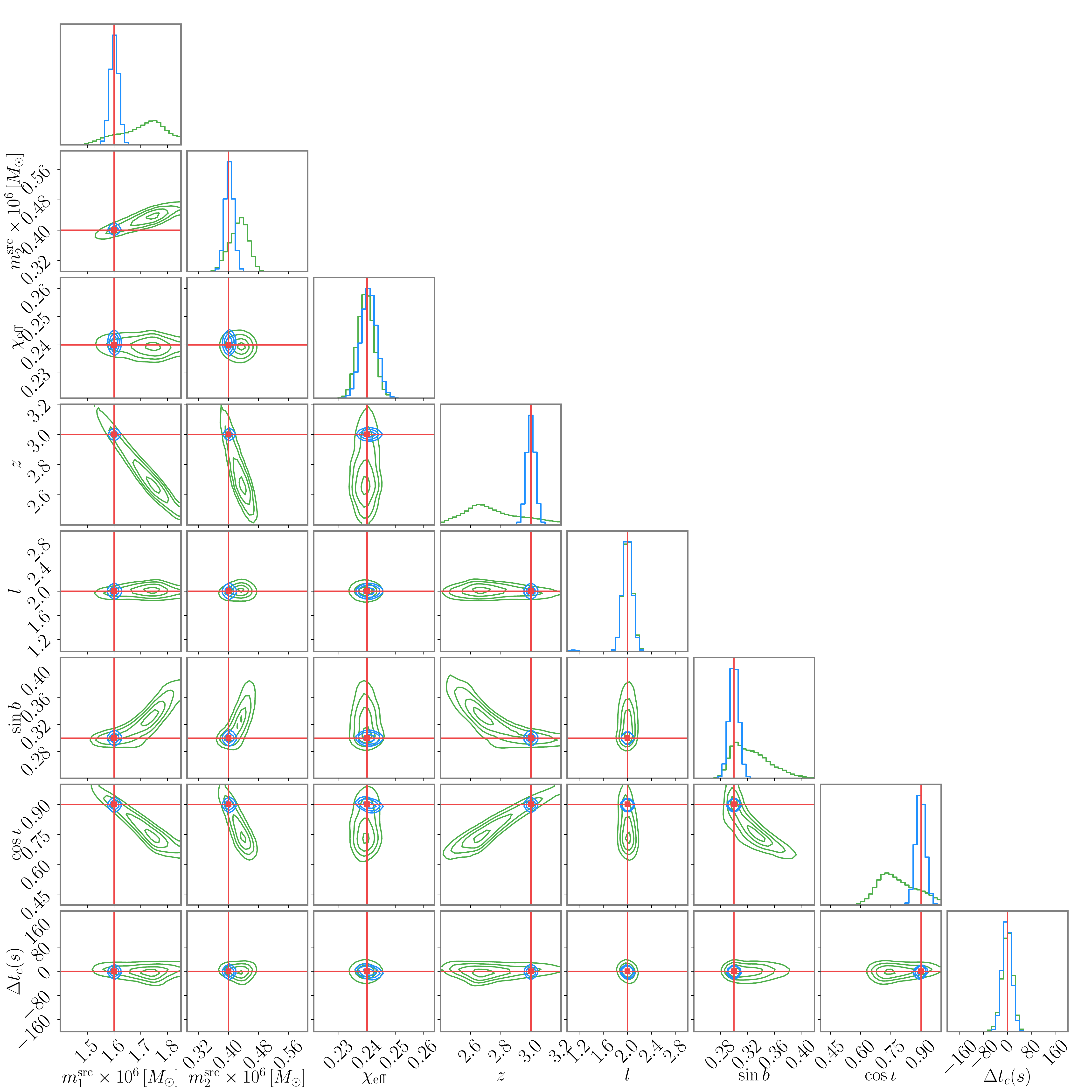}
        \caption{
        Binary III. Run with higher multipoles (blue) and with the $(2,2)$-mode only (green). The red lines denote the injected values. 
        }
        \label{fig:binary3_corner}
\end{figure*}

\begin{figure*}[ht!]
    \centering
    \includegraphics[width=\textwidth]{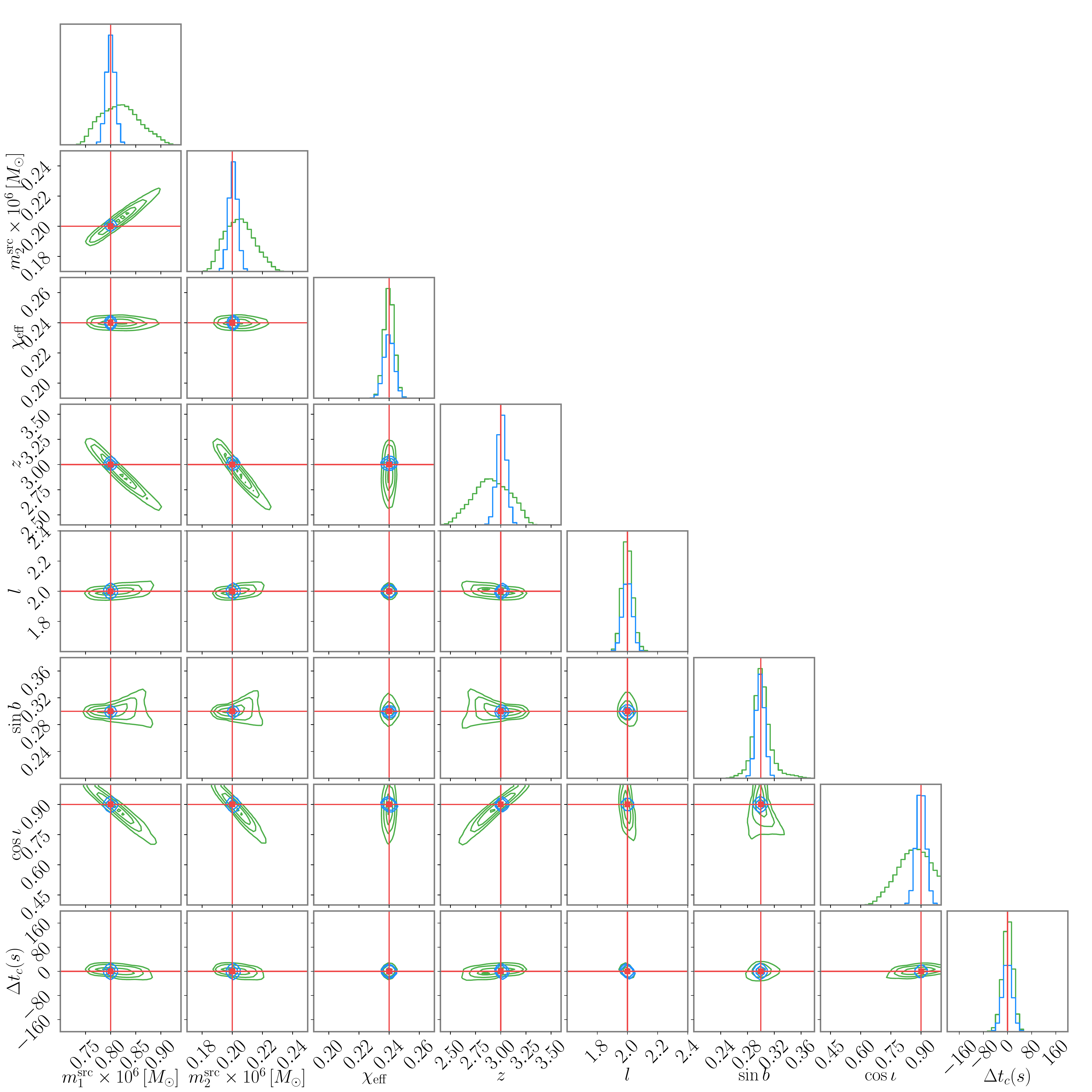}
        \caption{
        Binary IV. Run with higher multipoles (blue) and with the $(2,2)$-mode only (green). The red lines denote the injected values. 
        }
        \label{fig:binary4_corner}
\end{figure*}

\clearpage
\bibliographystyle{apsrev4-2}
\bibliography{lisa_mbhb.bib}

\end{document}